\documentclass[acus]{JAC2001} 

\usepackage{graphicx} 

\begin{document} 
\title{SOLITARY WAVES IN AN INTENSE BEAM PROPAGATING THROUGH A 
SMOOTH FOCUSING FIELD} 

\author{Stephan I. Tzenov\thanks{stzenov@pppl.gov} and Ronald C. 
Davidson \\ 
Plasma Physics Laboratory, Princeton University,
Princeton, New Jersey  08543} 

\maketitle 

\*\vspace{0.0cm}

\begin{abstract} 
Based on the Vlasov-Maxwell equations describing the self-consistent 
nonlinear beam dynamics and collective processes, the evolution of an 
intense sheet beam propagating through a periodic focusing field has 
been studied. In an earlier paper \cite{prstab} it has been shown 
that in the case of a beam with uniform phase space density the 
Vlasov-Maxwell equations can be replaced exactly by the macroscopic 
warm fluid-Maxwell equations with a triple adiabatic pressure law. 
In this paper we demonstrate that starting from the macroscopic 
fluid-Maxwell equations a nonlinear Schroedinger equation for the slowly 
varying wave amplitude (or a set of coupled nonlinear Schroedinger 
equations for the wave amplitudes in the case of multi-wave interactions) 
can be derived. Properties of the nonlinear Schroedinger equation are 
discussed, together with soliton formation in intense particle 
beams. 
\end{abstract} 

\renewcommand{\theequation}{\thesection.\arabic{equation}}

\setcounter{equation}{0}

\section{Introduction}

Of particular importance in modern accelerators and storage rings 
operating at high beam currents and charge densities are the effects 
of the intense self-fields produced by the beam space charge and 
current on determining detailed equilibrium, stability and transport 
properties. In general, a complete description of collective processes 
in intense charged particle beams is provided by the Vlasov-Maxwell 
equations for the self-consistent evolution of the beam distribution 
function and the electromagnetic fields. As shown in \cite{prstab} 
in the case of a sheet beam with constant phase-space density the 
Vlasov-Maxwell equations are fully equivalent to a warm-fluid model 
with zero heat flow and triple-adiabatic equation-of-state. 

In the present paper we demonstrate that starting from the hydrodynamic 
equations, and using the renormalization group (RG) technique 
\cite{cgoono,nozaki,oono,shiwa} a nonlinear Schroedinger equation for 
the slowly varying single-wave amplitude can be derived. The renormalized 
solution for the beam density describes the process of formation of 
periodic {\it holes} in intense particle beams. 

\renewcommand{\theequation}{\thesection.\arabic{equation}}

\setcounter{equation}{0}

\section{The Hydrodynamic Model}

We begin with the hydrodynamic model derived in \cite{prstab} 
\begin{eqnarray} 
{\frac {\partial \varrho} {\partial s}} + 
{\frac {\partial} {\partial x}} {\left( 
\varrho v \right)} = 0, 
\nonumber 
\end{eqnarray} 

\*\vspace{0.0cm}

\begin{equation} 
{\frac {\partial v} {\partial s}} + 
v {\frac {\partial v} {\partial x}} + 
v_T^2 {\frac {\partial} {\partial x}} 
\varrho^2 = - G(s) x - {\frac 
{\partial \psi} {\partial x}}, 
\label{Hydrodynamic} 
\end{equation} 
\begin{eqnarray} 
{\frac {\partial^2 \psi} {\partial x^2}} = 
- 2 \pi K \varrho. 
\nonumber 
\end{eqnarray} 
\noindent 
Here $\varrho {\left( x; s \right)} = n {\left( x; s \right)} / N$ 
and $v {\left( x; s \right)}$ are the normalized density and the 
current velocity, respectively, $G(s+S) = G(s)$ is the periodic 
focusing lattice coefficient, $v_T^2 = 3 {\widehat{P}}_0 / 2 
{\widehat{n}}_0^3$ is the normalized thermal speed-squared, and 
${\widehat{P}}_0 / {\widehat{n}}_0^3 = N^2 / 12 A^2$ is a constant 
coefficient \cite{prstab}, where $N$ is the area density of sheet 
beam particles, and $A$ is the constant phase-space density. Moreover, 
$\psi {\left( x; s \right)}$ is the normalized self-field 
potential 
\begin{eqnarray} 
\psi {\left( x; s \right)} = 
{\frac {e_b \phi {\left( x; s \right)}} 
{m_b \gamma_b \beta_b^2 c^2}}, 
\nonumber 
\end{eqnarray} 
\noindent 
where $\phi {\left( x; s \right)}$ is the electrostatic (space-charge) 
potential, $m_b$ and $e_b$ are the rest mass and charge of a beam 
particle, and $\beta_b$ and $\gamma_b$ are the relative particle 
velocity and Lorentz factor, respectively. Finally, the quantity 
$K$ is the normalized self-field perveance defined by 
\begin{eqnarray} 
K = {\frac {2N e_b^2} {m_b \gamma_b^3 
\beta_b^2 c^2}}. 
\nonumber 
\end{eqnarray} 

In what follows the analysis is restricted to the smooth focusing 
approximation 
\begin{equation} 
G(s) = G = {\rm const}, 
\label{Smooth} 
\end{equation} 
\noindent 
and assume that there exist nontrivial stationary solutions to 
(\ref{Hydrodynamic}) in the interval $x \in {\left( - x^{(-)}, 
x^{(+)} \right)}$, and that the sheet beam density is zero 
${\left( \varrho = 0 \right)}$ outside of the interval. The change 
of variables 
\begin{equation} 
\xi = x + x^{(-)}, 
\qquad \qquad 
\Psi = \psi - G x^{(-)} x 
\label{Change} 
\end{equation} 
\noindent 
enables us to rewrite (\ref{Hydrodynamic}) in the form 
\begin{eqnarray} 
{\frac {\partial \varrho} {\partial s}} + 
{\frac {\partial} {\partial \xi}} {\left( 
\varrho v \right)} = 0, 
\nonumber 
\end{eqnarray} 
\begin{equation} 
{\frac {\partial v} {\partial s}} + 
v {\frac {\partial v} {\partial \xi}} + 
v_T^2 {\frac {\partial} {\partial \xi}} 
\varrho^2 = - G \xi - {\frac 
{\partial \Psi} {\partial \xi}}, 
\label{Hydrodyn} 
\end{equation} 
\begin{eqnarray} 
{\frac {\partial^2 \Psi} {\partial \xi^2}} = 
- 2 \pi K \varrho. 
\nonumber 
\end{eqnarray} 
\noindent 
Clearly, the system (\ref{Hydrodyn}) possesses a stationary solution 
\begin{equation} 
\varrho_0 = {\frac {G} {2 \pi K}}, 
\quad \quad 
v_0 \equiv 0, \quad \quad 
\Psi_0 = - {\frac {G \xi^2} {2}} + 
{\rm const}. 
\label{Stationary} 
\end{equation} 
\noindent 
Here, the uniform density $\varrho_0$ is normalized according to 
\begin{equation} 
x^{(-)} + x^{(+)} = {\frac {1} {\varrho_0}} = 
{\frac {2 \pi K} {G}}. 
\label{Normalize} 
\end{equation} 
\noindent 

\renewcommand{\theequation}{\thesection.\arabic{equation}}

\setcounter{equation}{0}

\section{Renormalization Group Reduction of the Hydrodynamic Equations}

Following the basic idea of the RG method, we represent the solution 
to equations (\ref{Hydrodyn}) in the form of a standard perturbation 
expansion \cite{nayfeh} in a formal small parameter $\epsilon$ as 
\begin{equation} 
\varrho = \varrho_0 + \sum \limits_{k=1}^{\infty} 
\epsilon^k \varrho_k, \qquad \qquad 
v = \sum \limits_{k=1}^{\infty} 
\epsilon^k v_k, 
\label{Naiverhov} 
\end{equation} 
\begin{eqnarray} 
\Psi = - {\frac {G \xi^2} {2}} + 
\sum \limits_{k=1}^{\infty} 
\epsilon^k \Psi_k. 
\nonumber 
\end{eqnarray} 
\noindent 
Before proceeding with explicit calculations order by order, we note 
that in all orders the perturbation equations acquire the general 
form 
\begin{eqnarray} 
{\frac {\partial \varrho_n} {\partial s}} + 
\varrho_0 {\frac {\partial v_n} {\partial \xi}} 
= \alpha_n, 
\nonumber 
\end{eqnarray} 
\begin{equation} 
{\frac {\partial v_n} {\partial s}} + 
2 \varrho_0 v_T^2 {\frac {\partial \varrho_n} 
{\partial \xi}} = - {\frac {\partial \Psi_n} 
{\partial \xi}} + \beta_n, 
\label{Hydropert} 
\end{equation} 
\begin{eqnarray} 
{\frac {\partial^2 \Psi_n} {\partial \xi^2}} = 
- 2 \pi K \varrho_n, 
\nonumber 
\end{eqnarray} 
\noindent 
where the functions $\alpha_n {\left( \xi; s \right)}$ and  $\beta_n 
{\left( \xi; s \right)}$ involve contributions from previous orders 
and are considered known. Eliminating $v_n$ and $\Psi_n$, it is possible 
to obtain a single equation for $\varrho_n$ alone, i.e., 
\begin{equation} 
{\frac {\partial^2 \varrho_n} {\partial s^2}} - 
2 \varrho_0^2 v_T^2 {\frac {\partial^2 \varrho_n} 
{\partial \xi^2}} + G \varrho_n = 
{\frac {\partial \alpha_n} {\partial s}} - 
\varrho_0 {\frac {\partial \beta_n} {\partial \xi}}. 
\label{Basicrhon} 
\end{equation} 
\noindent 
It is evident that in first order $\alpha_1 = \beta_1 = 0$. Imposing 
the condition 
\begin{equation} 
\int \limits_{0}^{1/ \varrho_0} {\rm d} \xi 
\varrho_1 {\left( \xi; s \right)} = 0, 
\label{Conservation} 
\end{equation} 
\noindent 
which means that linear perturbation to the uniform stationary density 
$\varrho_0$ should average to zero and not affect the normalization 
properties on the interval ${\left( 0, x^{(-)} + x^{(+)} \right)}$, 
we obtain the first-order solution 
\begin{equation} 
\varrho_1 {\left( \xi; s \right)} = \sum 
\limits_{m \neq 0} {\cal A}_m e^{i \chi_m 
{\left( \xi; s \right)}}, \quad 
\chi_m {\left( \xi; s \right)} = 
\omega_m s + m \sigma \xi. 
\label{Solrho1} 
\end{equation} 
\noindent 
Here, ${\cal A}_m$ are constant complex wave amplitudes, and the 
following conventions and notations 
\begin{equation} 
\omega_{-m} = - \omega_m, \quad \quad 
\sigma = {\frac {G} {K}}, \quad \quad 
{\cal A}_{-m} = {\cal A}_m^{\ast}. 
\label{Notations} 
\end{equation} 
\noindent 
have been introduced. Moreover, the discrete mode frequencies $\omega_m$ 
are determined from the dispersion relation 
\begin{equation} 
\omega_m^2 = G + {\frac {v_T^2 \sigma^4} 
{2 \pi^2}} m^2. 
\label{Dispersion} 
\end{equation} 
\noindent 
In addition, the first-order solution for the current velocity and 
for the self-field potential can be expressed as 
\begin{equation} 
v_1 {\left( \xi; s \right)} = - {\frac {1} 
{\varrho_0 \sigma}} \sum \limits_{m \neq 0} 
{\frac {\omega_m} {m}} {\cal A}_m e^{i \chi_m 
{\left( \xi; s \right)}}, 
\label{Solv1} 
\end{equation} 
\begin{equation} 
\Psi_1 {\left( \xi; s \right)} = {\frac {2 \pi K} 
{\sigma^2}} \sum \limits_{m \neq 0} 
{\frac {{\cal A}_m} {m^2}} e^{i \chi_m 
{\left( \xi; s \right)}}. 
\label{Solpsi1} 
\end{equation} 

In obtaining the second-order perturbation equation (\ref{Basicrhon}), 
we note that 
\begin{equation} 
\alpha_2 = - {\frac {\partial} {\partial \xi}} 
{\left( \varrho_1 v_1 \right)}, \quad \quad 
\beta_2 = - {\frac {1} {2}} {\frac {\partial} 
{\partial \xi}} {\left( v_1^2 + 2 v_T^2 
\varrho_1^2 \right)}. 
\label{Alphabeta2} 
\end{equation} 
\noindent 
Thus the second-order solution for the density $\varrho_2 {\left( 
\xi; s \right)}$ is found to be 
\begin{equation} 
\rho_2 {\left( \xi; s \right)} = - \sum 
\limits_{m, k \neq 0} \alpha_{mk} {\cal A}_m 
{\cal A}_k e^{i {\left[ \chi_m {\left( 
\xi; s \right)} + \chi_k {\left( \xi; s 
\right)} \right]}}, 
\label{Solrho2} 
\end{equation} 
\noindent 
where 
\begin{eqnarray} 
\alpha_{mk} = {\frac {m + k} 
{{\cal D}_{mk}}} {\left[ {\frac {\omega_k 
{\left( \omega_m + \omega_k \right)}} 
{k \varrho_0}} \right.} 
\nonumber 
\end{eqnarray} 
\begin{equation} 
{\left. + {\frac {m + k} {2 \varrho_0}} 
{\left( {\frac {v_T^2 \sigma^4} {2 \pi^2}} + 
{\frac {\omega_m \omega_k} {mk}} 
\right)} \right]}, 
\label{Alphamkpm} 
\end{equation} 
\begin{equation} 
{\cal D}_{mk} = - {\left( \omega_m + \omega_k 
\right)}^2 + {\frac {v_T^2 \sigma^4} {2 \pi^2}} 
(m + k)^2 + G. 
\label{Dispmkpm} 
\end{equation} 
\noindent 
Having determined $\varrho_2$, the second-order current velocity 
$v_2 {\left( \xi; s \right)}$ can be found in a straightforward manner. 
The result is 
\begin{equation} 
v_2 {\left( \xi; s \right)} = {\frac {1} 
{\varrho_0 \sigma}} \sum \limits_{m, k \neq 0} 
\beta_{mk} {\cal A}_m {\cal A}_k e^{i {\left[ 
\chi_m {\left( \xi; s \right)} + 
\chi_k {\left( \xi; s \right)} \right]}}, 
\label{Solv2} 
\end{equation} 
\noindent 
where 
\begin{equation} 
\beta_{mk} = {\frac {\omega_k} 
{k \varrho_0}} + {\frac {\omega_m + 
\omega_k} {m + k}} \alpha_{mk}, 
\quad \quad 
\beta_{m, -m} = 0. 
\label{Betamkpm} 
\end{equation} 

In third order, the functions $\alpha_3$ and $\beta_3$ entering the 
right-hand-side of equation (\ref{Basicrhon}) can be calculated 
utilizing the already determined quantities from the first and second 
orders, according to 
\begin{equation} 
\alpha_3 = - {\frac {\partial} {\partial \xi}} 
{\left( \varrho_1 v_2 + \varrho_2 v_1 \right)}, 
\label{Alpha3} 
\end{equation} 
\begin{equation} 
\beta_3 = - {\frac {\partial} 
{\partial \xi}} {\left( v_1 v_2 + 
2 v_T^2 \varrho_1 \varrho_2 \right)}. 
\label{Beta3} 
\end{equation} 
\noindent 
It is important to note that the right-hand-side of equation 
(\ref{Basicrhon}) for $\varrho_3$ contains terms which yield 
oscillating terms with constant amplitudes to the solution for 
$\varrho_3$. Apart from these, there is a resonant term 
(proportional to $e^{i \chi_m {\left( \xi; s \right)}}$) leading 
to a secular contribution. To complete the renormalization group 
reduction of the hydrodynamic equations, we select this particular 
resonant third-order term on the right-hand-side of equation 
(\ref{Basicrhon}). The latter can be written as 
\begin{eqnarray} 
{\left( {\frac {\partial \alpha_3} {\partial s}} - 
\varrho_0 {\frac {\partial \beta_3} 
{\partial \xi}} \right)}_{res} = 
\nonumber 
\end{eqnarray} 
\begin{equation} 
\sum \limits_{m, k \neq 0} \Gamma_{mk} 
{\cal A}_m {\left| {\cal A}_k \right|}^2 
e^{i \chi_m {\left( \xi; s \right)}}, 
\label{Resrhs3} 
\end{equation} 
\noindent 
where 
\begin{eqnarray} 
\Gamma_{mk} = {\frac {m} {\varrho_0}} 
{\left[ \omega_m {\left( \beta_{mk} + {\frac 
{\omega_k \alpha_{mk}} {k}} \right)} \right.} 
\nonumber 
\end{eqnarray} 
\begin{equation} 
{\left. + {\frac {m \omega_k \beta_{mk}} {k}} 
+ {\frac {v_T^2 \sigma^4} {2 \pi^2}} m 
\alpha_{mk} \right]}. 
\label{Gammamk} 
\end{equation} 
\noindent 
Some straightforward algebra yields the solution for $\varrho_3 
{\left( \xi; s \right)}$ to equation (\ref{Basicrhon}) in the form 
\begin{equation} 
\varrho_3 {\left( \xi; s \right)} = \sum 
\limits_{m \neq 0} {\cal P}_m {\left( \xi; 
s \right)} e^{i \chi_m {\left( \xi; s 
\right)}} + \dots, 
\label{Solrho3} 
\end{equation} 
\noindent 
where the dots stand for non-secular oscillating terms. Moreover, the 
amplitude ${\cal P}_m {\left( \xi; s \right)}$ is secular and satisfies 
the equation 
\begin{equation} 
{\widehat{\cal L}}_m {\left( \xi; s 
\right)} {\cal P}_m {\left( \xi; s 
\right)} = \sum \limits_{k \neq 0} 
\Gamma_{mk} {\cal A}_m 
{\left| {\cal A}_k \right|}^2, 
\label{Equatpm} 
\end{equation} 
\noindent 
where the operator ${\widehat{\cal L}}_m$ is defined by 
\begin{equation} 
{\widehat{\cal L}}_m = {\frac {\partial^2} 
{\partial s^2}} + 2i {\left( \omega_m 
{\frac {\partial} {\partial s}} - 
{\frac {v_T^2 \sigma^3} {2 \pi^2}} m 
{\frac {\partial} {\partial \xi}} \right)} - 
{\frac {v_T^2 \sigma^2} {2 \pi^2}} 
{\frac {\partial^2} {\partial \xi^2}}. 
\label{Operatlm} 
\end{equation} 

We can now construct the perturbative solution for $\varrho$ up to 
third order in the small parameter $\epsilon$. Confining attention 
to the constant stationary density $\varrho_0$ and the fundamental 
modes (first harmonic in the phase $\chi_m$), we obtain 
\begin{equation}
\varrho {\left( \xi; s \right)} = \varrho_0 
+ \epsilon \sum \limits_{m \neq 0} {\left[ 
{\cal A}_m + \epsilon^2 {\cal P}_m 
{\left( \xi; s \right)} \right]} e^{i \chi_m 
{\left( \xi; s \right)}}.  
\label{Solrhogen} 
\end{equation} 
\noindent 
Following the basic philosophy of the RG method, we introduce the 
intermediate coordinate $X$ and ``time'' $S$ and transform equation 
(\ref{Solrhogen}) to 
\begin{eqnarray}
\varrho {\left( \xi; s \right)} = \varrho_0 
+ \epsilon \sum \limits_{m \neq 0} {\left\{ 
{\cal A}_m {\left( X; S \right)} \right.} 
\nonumber 
\end{eqnarray} 
\begin{equation}
{\left. + \epsilon^2 
{\left[ {\cal P}_m {\left( \xi; s \right)} - 
{\cal P}_m {\left( X; S \right)} \right]} 
\right\}} e^{i \chi_m {\left( \xi; s \right)}}.  
\label{Solrhoren} 
\end{equation} 
\noindent 
Note that the transition from equation (\ref{Solrhogen}) to equation 
(\ref{Solrhoren}) can always be performed by enforcing the constant 
amplitude ${\cal A}_m$ to be dependent on $X$ and $S$, which is in 
fact the procedure for renormalizing the standard perturbation result. 
Since the general solution for $\varrho {\left( \xi; s \right)}$ 
should not depend on $X$ and $S$, by applying the operator 
${\widehat{\cal L}}_m {\left( X; S \right)}$ [which is the same as 
that in equation (\ref{Operatlm}) but with $\xi \rightarrow X$ and 
$s \rightarrow S$] on both sides of equation (\ref{Solrhoren}), we 
obtain 
\begin{equation} 
{\widehat{\cal L}}_m {\left( X; S 
\right)} {\cal A}_m {\left( X; S 
\right)} = \sum \limits_{k \neq 0} 
\Gamma_{mk} {\cal A}_m {\left( 
X; S \right)} {\left| {\cal A}_k 
{\left( X; S \right)} \right|}^2, 
\label{Equatrgr} 
\end{equation} 
\noindent 
where we have dropped the formal parameter $\epsilon$ on the 
right-hand-side. Since the above equation should hold true for any 
choice of $X$ and $S$, we can set $X = \xi$ and $S = s$. Thus, we 
obtain the so-called proto RG equation \cite{nozaki,oono,shiwa}
\begin{equation} 
{\widehat{\cal L}}_m {\left( \xi; s 
\right)} {\cal A}_m {\left( \xi; s 
\right)} = \sum \limits_{k \neq 0} 
\Gamma_{mk} {\cal A}_m {\left( 
\xi; s \right)} {\left| {\cal A}_k 
{\left( \xi; s \right)} \right|}^2. 
\label{Equatprotorg} 
\end{equation} 
\noindent 
Introducing the new variable 
\begin{equation} 
\zeta_m = {\frac {v_T^2 \sigma^3 m} 
{2 \pi^2}} s + \omega_m \xi 
\label{Zetam} 
\end{equation} 
\noindent 
and neglecting the second order derivatives $\partial^2 / \partial 
s^2$ and $\partial^2 / \partial s \partial \zeta_m$, we finally 
arrive at the RG equation for the $m$-th mode amplitude 
\begin{equation} 
2i \omega_m {\frac {\partial {\cal A}_m} 
{\partial s}} - {\frac {v_T^2 \sigma^2 G} 
{2 \pi^2}} {\frac {\partial^2 {\cal A}_m} 
{\partial \zeta_m^2}} = \sum \limits_{k \neq 0} 
\Gamma_{mk} {\cal A}_m  {\left| {\cal A}_k 
\right|}^2. 
\label{Rgequation} 
\end{equation} 
\noindent 

\renewcommand{\theequation}{\thesection.\arabic{equation}}

\setcounter{equation}{0}

\section{The Nonlinear Schroedinger Equation for a Single Mode}

Equation (\ref{Rgequation}) represents a system of coupled nonlinear 
Schroedinger equations for the mode amplitudes. Neglecting the 
contribution from modes with $k \neq m$, for a single mode amplitude 
${\cal A}_m$, we obtain the equation 
\begin{equation} 
2i \omega_m {\frac {\partial {\cal A}_m} 
{\partial s}} - {\frac {v_T^2 \sigma^2 G} 
{2 \pi^2}} {\frac {\partial^2 {\cal A}_m} 
{\partial \zeta_m^2}} = - \Gamma_m {\left| 
{\cal A}_m \right|}^2 {\cal A}_m, 
\label{Nschroedeq} 
\end{equation} 
\noindent 
where 
\begin{equation} 
\Gamma_m = - \Gamma_{mm} = {\frac {2} 
{3G \varrho_0^2}} {\left( 16 \omega_m^4 - 
11G \omega_m^2 + G^2 \right)}. 
\label{Gammasch} 
\end{equation} 
\noindent 
It is easy to verify that $\Gamma_m$ is always positive. In nonlinear 
optics equation (\ref{Nschroedeq}) is known to describe the formation 
and evolution of the so-called {\it dark solitons} \cite{kivshar}. In 
the case of charged particle beams these correspond to the formation 
of {\it holes} or {\it cavitons} in the beam. Since the renormalized 
solution for the beam density $\varrho {\left( \xi; s \right)}$ can be 
expressed as 
\begin{equation}
\varrho {\left( \xi; s \right)} = \varrho_0 
+ \sum \limits_{m \neq 0} {\cal A}_m {\left( 
\xi; s \right)} e^{i \chi_m {\left( \xi; s \right)}}.  
\label{Rhorenorm} 
\end{equation} 
\noindent 
these holes have periodic structure in space $\xi$ and ``time'' $s$. 

\section{Concluding Remarks} 

Based on the renormalization group method, a system of coupled nonlinear 
Schroedinger equations has been derived for the slowly varying 
amplitudes of interacting beam-density waves. Under the approximation 
of an isolated wave neglecting the effect of the rest of the waves, 
this system reduces to a single nonlinear Schroedinger equation with 
repulsive nonlinearity. The latter describes the formation and 
evolution of holes in intense charged particle beams. 

\subsection*{Acknowledgments}

We are indebted to E. Startsev for many illuminating discussions 
concerning the subject of the present paper. It is also a pleasure 
to thank Y. Oono for careful reading of the manuscript and for making 
valuable suggestions. This research was supported by the U.S. 
Department of Energy under contract DE-AC02-76CH03073.

%%%%% Bibliography
%%

\end{document}